# PSYCHOLOGICAL DETERMINANTS OF ACADEMIC INTEGRITY IN THE USE OF GENERATIVE AI IN HIGHER EDUCATION

## ÜRETİCİ YAPAY ZEKÂNIN YÜKSEKÖĞRETİMDE *KULLANIMINDA AKADEMİK DÜRÜSTLÜĞÜN PSİKOLOJİK BELİRLEYİCİLERİ*

**Ezgi Dagtekin**[1]
**Ercan Erkalkan**[2]

**Abstract**

This paper examines the psychological determinants that shape academically honest and dishonest uses of generative artificial intelligence (GenAI) in higher education. Rather than treating academic misconduct as a purely technological problem, the study conceptualizes academic integrity as a psychologically mediated decision process influenced by moral reasoning, perceived social norms, policy clarity, academic self-efficacy, AI literacy, performance pressure, and beliefs about authorship. Methodologically, the paper adopts a focused narrative review and conceptual synthesis design. A purposive corpus of 16 core publications, including peer-reviewed studies and policy-oriented texts published between 2022 and March 2026, was assembled through targeted searches using combinations of the keywords generative AI, academic integrity, academic misconduct, moral disengagement, AI literacy, and higher education. The reviewed literature suggests that students do not interpret all forms of AI assistance as cheating. Integrity risk increases when institutional guidance is vague, peer use appears normalized, academic pressure is high, and AI tools are perceived as legitimate substitutes for difficult cognitive labor. By contrast, assignment-level guidance, explicit disclosure norms, ethics-oriented instruction, and authentic assessment design appear to reduce integrity risk more effectively than detection-centered responses alone. Based on these findings, the paper proposes an integrative conceptual model in which institutional context shapes psychological appraisal, and psychological appraisal in turn influences disclosed, borderline, or dishonest GenAI use. The paper concludes that effective responses to GenAI-related integrity problems should combine policy clarity, pedagogy, AI literacy, and student support rather than relying only on prohibition or software-based surveillance.

**Keywords:** academic integrity, generative AI, higher education, moral disengagement, AI literacy

[1] The MacDuffie School, Massachusetts, USA

[2] Dr. Öğr. Üyesi, Department of Electronics and Automation, Marmara University, ercan.erkalkan@marmara.edu.tr, https://orcid.org/0000-0001-9259-7112)

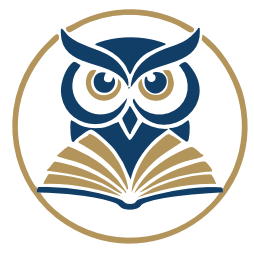

**Özet**

Bu bildiri, üretici yapay zekânın (ÜYZ) yükseköğretimde akademik olarak dürüst ve dürüst olmayan kullanım biçimlerini şekillendiren psikolojik belirleyicileri incelemektedir. Çalışma, akademik usulsüzlüğü yalnızca teknolojik bir sorun olarak ele almak yerine, akademik dürüstlüğü ahlaki muhakeme, algılanan sosyal normlar, politika açıklığı, akademik öz-yeterlik, yapay zekâ okuryazarlığı, performans baskısı ve yazarlık algısından etkilenen psikolojik olarak aracılanmış bir karar süreci olarak kavramsallaştırmaktadır. Yöntemsel olarak çalışma, odaklı anlatısal literatür incelemesi ve kavramsal sentez tasarımını benimsemektedir. 2022 ile Mart 2026 arasında yayımlanan hakemli çalışmalar ile politika odaklı metinlerden oluşan ve üretici yapay zekâ, akademik dürüstlük, akademik usulsüzlük, ahlaki çözülme, yapay zekâ okuryazarlığı ve yükseköğretim anahtar sözcükleriyle taranan amaçlı bir 16 kaynaklık derlem oluşturulmuştur. İncelenen literatür, öğrencilerin tüm yapay zekâ destek biçimlerini otomatik olarak kopya olarak değerlendirmediklerini göstermektedir. Kurumsal rehberliğin belirsiz olduğu, akran kullanımının normalleştiği, akademik baskının yükseldiği ve ÜYZ araçlarının zorlayıcı bilişsel emeğin meşru bir ikamesi olarak algılandığı durumlarda dürüstlük riski artmaktadır. Buna karşılık ödev düzeyinde açık kullanım kuralları, net beyan normları, etik odaklı öğretim ve özgün ölçme-değerlendirme tasarımları, yalnızca tespit odaklı yaklaşımlardan daha etkili görünmektedir. Bu bulgulara dayanarak çalışma, kurumsal bağlamın psikolojik değerlendirmeyi, psikolojik değerlendirmenin ise beyan edilen, sınırda kalan veya dürüst olmayan ÜYZ kullanımını etkilediği bütünleştirici bir kavramsal model önermektedir. Sonuç olarak, ÜYZ kaynaklı akademik dürüstlük sorunlarına verilecek etkili yanıtların yalnızca yasaklama veya yazılım tabanlı gözetim yerine politika açıklığı, pedagojik yönlendirme, yapay zekâ okuryazarlığı ve öğrenci desteğini birlikte içermesi gerektiği savunulmaktadır.

## INTRODUCTION

Generative artificial intelligence has disrupted long-standing assumptions about authorship, originality, and acceptable assistance in higher education. Large language models can summarize readings, propose structures, paraphrase drafts, generate code, emulate voice, and even fabricate references in ways that blur the boundary between support and substitution. This shift has intensified academic integrity debates because the core issue is no longer limited to plagiarism in its traditional copy-and-paste form. Instead, the concern extends to hidden delegation of cognitive work, opaque collaboration with non-human agents, and the weakening of students' ownership of their submitted work (Perkins, 2023; Roe & Perkins, 2022; Duah & McGivern, 2024).

Recent empirical studies show that students interpret the integrity status of GenAI use unevenly. In an Australian survey, more than one third of students reported using a chatbot to assist with assessment, and many did not automatically view that behavior as an academic integrity breach (Gruenhagen et al., 2024). A broader multicultural study spanning 76 countries similarly found widespread awareness and familiarity with GenAI, but considerable variation in how participants evaluated its legitimacy in academic work (Yusuf et al., 2024). These findings suggest that integrity judgments are not solely determined by formal rules. They are filtered through beliefs about fairness, effort, pressure, and what peers appear to do.

For this reason, an exclusively technological or punitive framing is insufficient. Detection tools, authorship checks, and stricter sanctions may still have a role, but they do not fully explain why students choose disclosure, borderline reliance, or dishonest substitution in the first place. Studies on GenAI-related cheating increasingly point to psychological processes such as moral disengagement, subjective norms, rationalization, perceived utility, and self-efficacy (Zhang et al., 2024; Huang et al., 2025; Theoharakis et al., 2025). Moreover, ambiguity in institutional guidance can function as a silent permissive signal, especially when classroom expectations and university-level policy are misaligned (Gonsalves, 2025; Huang et al., 2025).

The present paper addresses a practically significant question: which psychological determinants most strongly shape academic integrity in the use of GenAI in higher education? The paper makes three contributions. First, it synthesizes recent literature on the psychological drivers of GenAI-related integrity decisions. Second, it proposes an integrative conceptual model linking institutional context, psychological appraisal, and behavioral outcomes. Third, it derives practical implications for policy design, pedagogy, disclosure practices, and assessment design. The central argument is that academic integrity in GenAI use should be understood as a situated decision process in which students evaluate moral cost, perceived normality, personal competence, performance pressure, and the legitimacy of AI assistance before acting.

## 1. METHOD

This study uses a focused narrative review and conceptual synthesis design. The aim is not to produce a formal systematic review or meta-analysis, but to identify and interpret the psychological determinants that recur across recent literature on academic integrity in the use of GenAI in higher education. This design was selected because the field is rapidly evolving, the available evidence is methodologically heterogeneous, and many relevant publications combine empirical findings with policy interpretation.

### 1.1. Conceptual Framework

The analysis is grounded in a behavioral ethics perspective. Two complementary theoretical lenses guide the synthesis. First, the Theory of Planned Behavior explains how attitudes, subjective norms, and perceived behavioral control shape intention (Ajzen, 1991). This lens is appropriate because many GenAI-related academic decisions involve anticipatory judgments about whether a specific AI-supported action will be considered acceptable. Second, moral disengagement theory explains how individuals reduce the perceived ethical cost of questionable behavior by minimizing harm, diffusing responsibility, or redefining misconduct as harmless, necessary, or normal (Bandura, 1999). Together, these perspectives support an interpretation of GenAI misuse as a psychologically mediated and context-sensitive decision rather than a simple rule violation.

### 1.2. Review Procedure

The literature corpus was developed through targeted searches conducted on Google Scholar, Scopus-indexed journal portals accessible through institutional search, and publisher databases commonly used in education and technology research. Searches were conducted iteratively and finalized in March 2026. Search combinations included the terms generative AI, ChatGPT, academic integrity, academic misconduct, cheating, moral disengagement, AI literacy, authorship, disclosure, and higher education, used in combinations such as “generative AI” AND “academic integrity”, “ChatGPT” AND cheating, and “AI literacy” AND higher education. The temporal scope was limited to publications from 2022 to March 2026 because GenAI became a mainstream issue in higher education during this period. Initial screening prioritized peer-reviewed journal articles, conference papers, and policy-oriented reports with direct relevance to integrity, disclosure, acceptable use, policy ambiguity, or psychological drivers of AI use in academic settings.

### 1.3. Inclusion and Exclusion Criteria

Publications were included if they met three conditions: first, they addressed GenAI use in higher education or closely related academic settings; second, they contained either empirical findings, conceptual analysis, or policy discussion directly relevant to academic integrity; third, they examined at least one psychological, social, or behavioral determinant of AI-related academic conduct. Commentary texts without analytical substance, duplicate records, and publications focused solely on technical model performance without educational or ethical implications were excluded.

### 1.4. Corpus and Analytical Strategy

The final review corpus consisted of 16 core review texts, including 13 peer-reviewed empirical or conceptual studies and 3 policy or guidance texts. Foundational theoretical sources used to frame the analysis, such as Ajzen (1991), Bandura (1999), and Long and Magerko (2020), informed the conceptual background but were not counted among the 16 core review texts. The sample was purposive rather than random because the objective was conceptual relevance rather than statistical representation. Because the study was designed as a focused narrative review rather than a PRISMA-style systematic review, the search and selection process was iterative and did not rely on a formal record-by-record screening flow with exhaustive counts. Instead, texts were retained until the recurring determinants and intervention themes were sufficiently covered for conceptual synthesis. After full-text reading, each document was coded for recurring determinants such as moral disengagement, perceived normativity, policy clarity, academic self-efficacy, AI literacy, time pressure, and authorship beliefs. In a second coding round, the direction of influence of each factor was examined to determine whether it appeared to increase integrity risk, reduce it, or operate conditionally. The synthesis then grouped the recurring determinants into a conceptual model linking institutional context, psychological appraisal, and behavioral outcome.

## 2. REVIEW SCOPE AND ANALYTICAL BASIS

Before presenting the substantive determinants in Section 3, this section defines the analytical scope of the review and clarifies how the selected literature was treated as the basis for the conceptual synthesis.

### 2.1. Purpose and Significance of the Study

The main purpose of the study is to identify the psychological conditions under which GenAI use is more likely to remain academically legitimate or shift toward dishonest reliance. The significance of the topic lies in the fact that institutional responses to GenAI still tend to focus on detection, prohibition, or generic rule statements, whereas recent research increasingly indicates that students' conduct is shaped by how they interpret the fairness, meaning, legitimacy, and risks of AI use. A psychologically informed account is therefore necessary to explain why the same tool may be used by one student for planning, revision, and language support, yet by another for undisclosed substitution of authorship.

### 2.2. Review Corpus

The review corpus consists of 16 core texts, including 13 peer-reviewed empirical or conceptual studies and 3 policy or guidance documents. The corpus was constructed purposively to capture methodological variation while preserving close relevance to higher education and academic integrity. The selected texts jointly represent student-centered evidence, policy interpretation, and pedagogical discussion, thereby allowing institutional and individual-level influences to be analyzed together.

### 2.3. Analytical Strategy

A thematic synthesis strategy was employed. Each document was coded for the presence of specific determinants, including moral disengagement, perceived social normativity, academic self-efficacy, AI literacy, performance pressure, policy awareness, and authorship beliefs. A second analytical round focused on whether each factor appeared to increase integrity risk, reduce it, or exert a conditional effect depending on institutional guidance or assessment design. Priority was given to determinants that recurred across independent sources or that were explicitly linked to behavioral outcomes such as disclosure, covert use, cheating intention, or normalization of AI-assisted work.

## 3. SYNTHESIZED FINDINGS

The reviewed literature converges on a set of interrelated determinants that shape how students interpret and use GenAI in academic work. These determinants do not operate independently. Instead, they interact through a broader process of psychological appraisal in which institutional context, perceived norms, and personal coping needs influence whether GenAI is used transparently, ambiguously, or dishonestly.

A first recurring determinant is moral disengagement. Studies on GenAI-related misconduct show that students frequently distinguish between serious cheating and more subtle AI-mediated practices, such as undisclosed paraphrasing, rewriting, or idea generation. This gray-zone reasoning allows misconduct to be reframed as efficiency, harmless support, or normal adaptation to a new technological environment. Zhang et al. (2024) show that moral and rational evaluations interact in shaping ChatGPT-based misconduct, while Theoharakis et al. (2025) specifically connect AI engagement to morally disengaged forms of misconduct. Huang et al. (2025) similarly report that personal ethics are central in students' decisions about academic cheating with GenAI.

A second determinant is social influence. GenAI use becomes easier to justify when students believe that everyone is doing it, that instructors tacitly expect AI fluency, or that peer competition rewards speed over process. Ajzen's (1991) framework predicts this pattern, and recent studies confirm it in AI contexts. Huang et al. (2025) found peer influence to be strongly associated with cheating behavior, while Acosta-Enriquez et al. (2025) showed that social influence and integrity perceptions are central to AI uptake in academic settings. The significance of this determinant is that institutions may unintentionally normalize misconduct through silence.

A third determinant is policy clarity and disclosure culture. Research consistently indicates that students do not automatically map formal university rules onto concrete GenAI practices. Duah and McGivern (2024) describe ambiguity around authorial identity and acceptable use, and Gonsalves (2025) shows that non-compliance in AI use declarations is not always straightforward defiance; it may also reflect confusion, inconsistent assessment design, or uncertainty about what exactly counts as reportable use. Importantly, Huang et al. (2025) found classroom-level policies to exert stronger influence than general institutional policies. This interpretation is consistent with broader discussions emphasizing that operational guidance, not abstract policy language alone, is needed to preserve academic integrity under GenAI conditions (Lund et al., 2025; TEQSA, 2024).

A fourth determinant concerns academic self-efficacy and AI literacy. Students with weak confidence in their ability to start, organize, or complete demanding tasks may treat GenAI as a substitute for effort rather than as support for learning. At the same time, low AI literacy can produce a different risk: students may not understand the limits, biases, or authorship implications of generated content. Long and Magerko (2020) define AI literacy as a set of competencies involving understanding, evaluation, and ethical use. Hill and Hargis (2024) demonstrate that structured ethics instruction can strengthen reflective engagement with these issues.

A fifth determinant is performance pressure and time scarcity. GenAI is especially attractive under conditions of deadline compression, grade anxiety, language insecurity, or excessive workload. Gruenhagen et al. (2024) identify psychosocial conditions associated with chatbot use in assessments, while Yusuf et al. (2024) show that participants across cultures simultaneously recognize the productivity benefits and integrity risks of GenAI. In such contexts, misuse may be driven less by explicit intent to deceive than by perceived necessity.

A final determinant is authorship belief. Academic integrity in GenAI use depends heavily on what students think ownership of work actually means. If authorship is interpreted narrowly as having submitted the file rather than having done the central intellectual work, then extensive undisclosed AI assistance can be framed as acceptable. Perkins (2023) and Bittle and El-Gayar (2025) both emphasize that GenAI challenges inherited definitions of plagiarism and originality. The key integrity question is therefore not simply whether AI was used, but whether the student remains epistemically and authorially responsible for the submitted product.

Across the corpus, moral disengagement, social influence, and policy clarity emerged as the most recurrent and behaviorally proximate determinants. Academic self-efficacy and performance pressure also appeared frequently, typically as enabling conditions that made overreliance on GenAI more likely under stress. By contrast, AI literacy and authorship beliefs were analytically important but less consistently operationalized in empirical terms, which suggests that these two determinants remain especially important targets for future measurement-focused research.

Table 1 summarizes the main psychological determinants identified in the reviewed literature, the mechanisms through which they operate, the integrity risks they create, and the corresponding educational responses.

**Table 1.** Psychological determinants of academic integrity in GenAI use

| Determinant | Psychological mechanism | Typical integrity risk | Suggested educational response |
|---|---|---|---|
| Moral disengagement | Redefining misconduct as harmless efficiency | Undisclosed substitution or paraphrasing | Ethics-focused case analysis and reflective prompts |
| Peer influence | Normalizing AI use through social comparison | Group-level normalization of covert use | Visible norms, task-specific examples, and classroom dialogue |
| Policy ambiguity | Uncertainty about what must be disclosed | Non-disclosure and boundary confusion | Assignment-level AI guidance and disclosure templates |
| Low academic self-efficacy | Delegating difficult cognitive work to the tool | Overreliance on generated drafts | Scaffolded drafting support and feedback checkpoints |
| Weak AI literacy | Inability to judge limitations, bias, or authorship implications | Fabricated references and shallow acceptance of outputs | AI literacy instruction tied to academic writing |
| Performance pressure | Using GenAI to cope with time, workload, or grade anxiety | Deadline-driven misuse and rationalization | Workload calibration and staged assessments |
| Authorship erosion | Submission treated as equivalent to authorship | Loss of ownership of intellectual work | Explicit teaching of authorship and epistemic responsibility |

Source: Prepared by the author based on the reviewed literature.

As Figure 1 illustrates, the findings support an integrative conceptual model consisting of three layers. At the first layer, institutional context includes policy clarity, assessment design, disclosure expectations, and visible classroom norms. At the second layer, psychological appraisal includes moral disengagement, perceived normativity, self-efficacy, AI literacy, performance pressure, and authorship beliefs. At the third layer, behavioral outcomes include disclosed support use, borderline reliance, and dishonest substitution. In this model, institutional context does not influence behavior directly in a simple linear fashion; rather, it shapes how students interpret the legitimacy, necessity, and ethical cost of GenAI use.

**Figure 1.** Integrative conceptual model of academic integrity in the use of GenAI

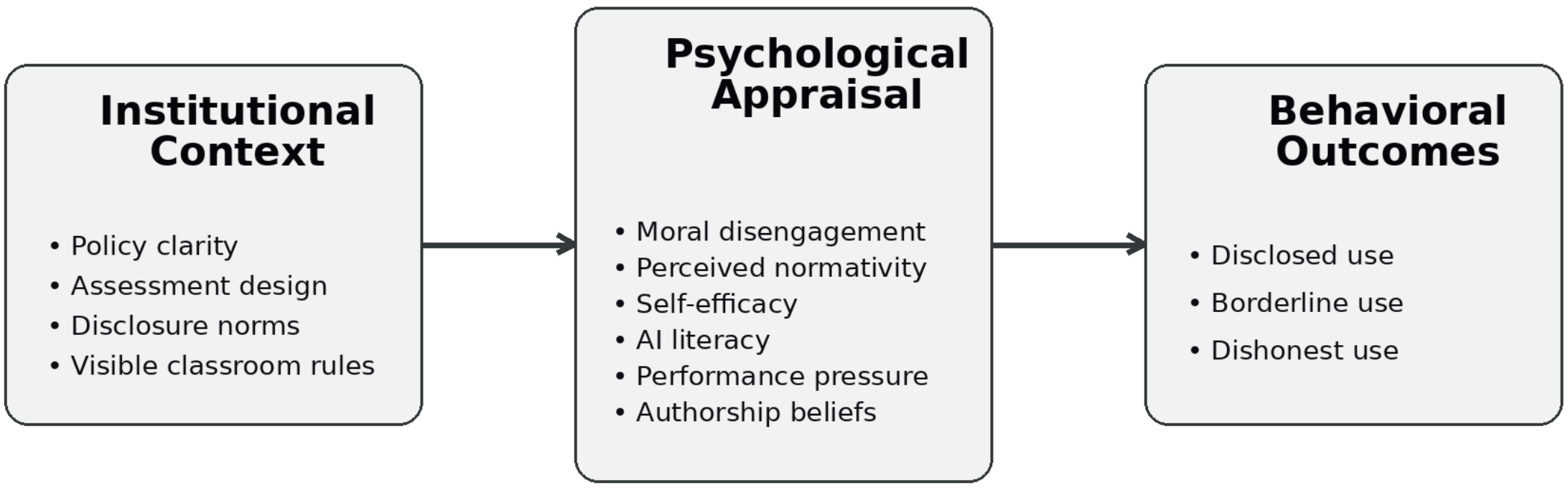

**Source:** Prepared by the author based on the reviewed literature.

## DISCUSSION AND CONCLUSION

The synthesis indicates that academic integrity in the use of GenAI is best understood as the outcome of an interaction between institutional context and psychological appraisal. Students do not merely respond to formal rules. They interpret assignment requirements, observe peers, estimate reputational and academic risk, compare effort with reward, and construct a justification for action. Under conditions of policy ambiguity, weak disclosure culture, high workload, and output-centered assessment, psychologically convenient rationalizations become more available, thereby increasing the likelihood of hidden or excessive GenAI reliance.

These findings have three direct implications. First, policy should move from generic prohibition to use-specific guidance. Broad statements that AI may not be used inappropriately are too vague to regulate psychologically ambiguous behaviors such as idea generation, editing, translation, summarization, or code debugging. Second, pedagogy should explicitly teach authorship, disclosure, and epistemic responsibility. Students need examples of acceptable, questionable, and unacceptable uses rather than abstract integrity slogans. Third, integrity work should be paired with AI literacy and academic support. A student who lacks confidence, time, or language resources is more likely to overdelegate to GenAI. This broader response also aligns with holistic framework proposals that combine policy, pedagogy, and student support rather than detection alone (Rasul et al., 2024).

The paper also suggests that integrity research in the GenAI era should widen its methodological focus. Future empirical work could test the proposed model by measuring how moral disengagement, authorship belief, AI literacy, and time pressure jointly predict disclosed and undisclosed AI use across disciplines. Experimental studies could compare assessment designs with and without process-based disclosure mechanisms. Cross-cultural work is also needed, because the review indicates that norms of assistance, collaboration, and fairness vary across institutional settings.

Taken together, the paper contributes in three ways: it synthesizes the recurring psychological determinants of GenAI-related integrity decisions, proposes a three-layer conceptual model linking institutional context, psychological appraisal, and behavioral outcomes, and translates that synthesis into practical implications for policy, pedagogy, disclosure, and assessment design.

The present study has several limitations. It is a focused narrative review rather than a formal systematic review or meta-analysis, and the final corpus was purposively assembled rather than statistically sampled. The evidence base is also still evolving, and some included publications are conceptual or policy-oriented rather than fully empirical. In addition, the review emphasizes higher education and may not fully capture integrity dynamics in secondary education or professional certification contexts. Nevertheless, the study offers a timely conceptual synthesis for a rapidly changing field and identifies clear directions for empirical testing.